\documentclass[conference]{IEEEtran}
\IEEEoverridecommandlockouts

\usepackage{cite}
\usepackage{amsmath,amssymb,amsfonts}
\usepackage{algorithmic}
\usepackage{graphicx}
\usepackage{textcomp}
\usepackage{xcolor}
\usepackage{booktabs}
\usepackage{multirow}
\usepackage{tikz}
\usetikzlibrary{positioning,shapes.geometric,shapes.symbols,arrows.meta,calc}

\def\BibTeX{{\rm B\kern-.05em{\sc i\kern-.025em b}\kern-.08em
    T\kern-.1667em\lower.7ex\hbox{E}\kern-.125emX}}
\usepackage[hidelinks]{hyperref}
\begin{document}

\title{Human Activity Recognition using Ceiling-Mounted Multi-Radar Sensors}

\author{\IEEEauthorblockN{Ananya Prabhu}
\IEEEauthorblockA{\textit{Faculty of ICT, Tampere University} \\
\textit{Pori University Consortium}\\
Pori, Finland \\
ananya.prabhu@tuni.fi}
\and
\IEEEauthorblockN{Aditi Site}
\IEEEauthorblockA{\textit{Faculty of ICT, Tampere University} \\
\textit{Pori University Consortium}\\
Pori, Finland \\
aditi.site@tuni.fi}

\and
\IEEEauthorblockN{Tarmo Lipping}
\IEEEauthorblockA{\textit{Faculty of ICT, Tampere University} \\
\textit{Pori University Consortium}\\
Pori, Finland \\
tarmo.lipping@tuni.fi}
}
\maketitle

\begin{abstract}
Continuous activity monitoring in assisted living should work without cameras or wearable devices. Radar meets this requirement, but most radar studies use a single sensor and activities recorded in isolation, and few deliver results in real time. We present an indoor monitoring system in which point clouds from three ceiling-mounted radar sensors are fused using density-based clustering, movement and posture features are derived for activity classification, and results are shown on a real-time dashboard. In a preliminary study with eight participants performing everyday activities as one continuous scenario, a random forest reached 74\% frame-level accuracy with a random 80/20 split and 68\% with leave-one-participant-out evaluation. Without position features, accuracy fell to 46\% and 44\%, indicating partial dependence on room layout, while window-level features without position reached 67\% and 65\%.
\end{abstract}

                   
\section{Introduction}
Human Activity Recognition (HAR) has been studied across various sensing modalities, including vision, inertial, and radar signals, along with machine learning (ML) and deep learning (DL) methods \cite{Shin25}. Wearable sensors such as accelerometers and gyroscopes provide accurate motion measurements, however, the effectiveness depends on the user's consistency in wearing them. Vision-based sensors avoid this dependency but compromise privacy conditions \cite{Shen24}. Radar in this case is contactless and privacy-preserving. It has been utilized for vital sign measurement, fall detection, gait analysis, and activity recognition \cite{Ahmed24}. Furthermore, unlike vision-based sensing, radar is effectively more reliable under poor illumination and fog \cite{Papadopoulos23}. This paper presents a preliminary study of an indoor monitoring system that fuses three ceiling-mounted radars, which recognizes daily activities and displays the results on a real time dashboard.

\section{Related Work}

\subsection{Radar-Based Human Activity Recognition}

HAR studies mostly implement single-activity recognition with fixed
start and end points under controlled laboratory conditions.
Continuous recognition, where activities have unknown durations and
arbitrary transitions, is relatively underexplored, even though this
is how activities occur in real deployments \cite{Ullmann23}.
Shrestha et al. \cite{Shrestha20} addressed this by applying
bidirectional long short-term memory (Bi-LSTM) networks to continuous
sequences of activities, achieving over 90\% mean accuracy with
micro-Doppler data compared with about 76\% with range-time data. Li
et al. \cite{Li20} combined frequency-modulated continuous-wave (FMCW)
radar with wearable inertial sensors and achieved about 96\% accuracy,
validated with leave-one-participant-out (LOPO) evaluation.

Radar data for HAR is widely represented as micro-Doppler
spectrograms, range-time maps, or range-Doppler maps. DL models can
learn features directly from these representations, in contrast to
classical ML methods applied to hand-crafted features, which have a
lower computational cost \cite{Papadopoulos23}. In practice, both are
often combined. Abdu et al. \cite{Abdu22} extracted micro-Doppler
features with pretrained convolutional neural networks (CNNs) and
performed classification using a support vector machine (SVM),
reaching around 99\% accuracy on activity and fall classification. CNN and SVM
are the most widely used models for activity recognition in
healthcare radar studies \cite{Ahmed24}.

\subsection{Fall Detection and Multi-Radar Tracking}

Fall detection is one of the most studied radar healthcare
applications. A survey on radar-based fall detection found that
traditional classifiers such as SVM and k-nearest neighbors (k-NN)
perform well, whereas DL models, specifically CNNs and recurrent
neural networks (RNNs), handle complex features and large
unstructured datasets better \cite{Hu24}. Li et al. \cite{Li24}
combined a CNN with a Bi-LSTM in a convolutional bidirectional LSTM
(CB-LSTM) model that captures spatial and temporal features from the
radar spectrogram, achieving about 98\% fall detection accuracy. Liang et al. \cite{Liang24}
observed that low-quality point clouds reduce the reliability of fall
detection and proposed a point cloud enhancement model for 24~GHz
FMCW radar, which achieved 99\% and 98\% accuracy on new participants
and new environments, respectively.

These systems mostly rely on a single radar. Tracking several people
with millimeter-wave (mmWave) radar and keeping their identities
across an interrupted field of view is still an open challenge, and
sensor fusion is identified as a key direction for progress
\cite{Pearce23}. Shen et al. \cite{Shen24} combined three mmWave
radars in a real-time framework using density-based spatial
clustering of applications with noise (DBSCAN), with parameters
adapted to the signal-to-noise ratio (SNR), and a probability matrix
for target tracking. Their system achieved tracking precision of
98\%, 96\%, and 94\% for one, two, and three people, and 96\% fall
detection accuracy. This study is the closest design to ours;
however, it does not extend to activity analysis or an interface.

\subsection{Long-Term Monitoring Systems}

Radar can further support long-term health assessment, where changes
in movement and daily routine provide useful information. Doppler radar has enabled contactless in-home gait analysis \cite{Seifert21}, and ultra-wideband radar has detected abnormal events from body movement and cardiorespiratory signals \cite{Diraco17}. Abedi et al. \cite{Abedi23} deployed a cloud-based FMCW radar system that monitors gait and activity and reports them to
nurses through dashboards. However, most clinician-oriented systems do not use radar at all. Shahid et al.
\cite{Shahid22} used Internet of Things (IoT) sensors to learn older
adults' daily routines and notify relatives of deviations, and users
rated the notifications positively.

\begin{table}[t]
\centering
\caption{Comparison of Existing Studies and the Proposed Framework}
\label{tab:comparison}
\scriptsize
\setlength{\tabcolsep}{4pt}
\begin{tabular}{lcccccccccc}
\toprule
\textbf{Ref.} & \textbf{MR} & \textbf{MT} & \textbf{CH} & \textbf{ML} & \textbf{AR} & \textbf{FD} & \textbf{AD} & \textbf{LT} & \textbf{CS} & \textbf{IF} \\
\midrule
\cite{Shin25}         & -- & -- & \checkmark & \checkmark & \checkmark & -- & -- & -- & -- & -- \\
\cite{Ullmann23}      & -- & -- & \checkmark & \checkmark & \checkmark & -- & -- & \checkmark & -- & $\circ$ \\
\cite{Ahmed24}        & -- & -- & \checkmark & \checkmark & \checkmark & \checkmark & -- & \checkmark & -- & $\circ$ \\
\cite{Miazek24}       & -- & -- & \checkmark & \checkmark & \checkmark & \checkmark & -- & \checkmark & -- & $\circ$ \\
\cite{Papadopoulos23} & -- & -- & \checkmark & \checkmark & \checkmark & \checkmark & -- & -- & -- & -- \\
\cite{Shrestha20}     & -- & -- & \checkmark & \checkmark & \checkmark & -- & -- & -- & -- & -- \\
\cite{Li20}           & -- & \checkmark & \checkmark & \checkmark & \checkmark & \checkmark & -- & -- & -- & $\circ$ \\
\cite{Abdu22}         & -- & -- & -- & \checkmark & \checkmark & -- & -- & -- & -- & -- \\
\cite{Zhao22}         & -- & -- & -- & \checkmark & \checkmark & -- & -- & \checkmark & -- & -- \\
\cite{Li24}           & -- & -- & -- & \checkmark & -- & \checkmark & -- & -- & -- & -- \\
\cite{Liang24}        & -- & -- & -- & \checkmark & -- & \checkmark & -- & -- & -- & -- \\
\cite{Shen24}         & \checkmark & \checkmark & -- & \checkmark & -- & \checkmark & -- & -- & -- & \checkmark \\
\cite{Pearce23}       & \checkmark & \checkmark & -- & \checkmark & -- & -- & -- & -- & -- & $\circ$ \\
\cite{Hu24}           & -- & -- & -- & \checkmark & -- & \checkmark & -- & -- & -- & -- \\
\cite{Diraco17}       & -- & -- & \checkmark & \checkmark & \checkmark & -- & \checkmark & \checkmark & -- & $\circ$ \\
\cite{Abedi23}        & -- & \checkmark & \checkmark & \checkmark & \checkmark & -- & -- & \checkmark & \checkmark & \checkmark \\
\cite{Shahid22}       & -- & -- & \checkmark & \checkmark & \checkmark & -- & \checkmark & \checkmark & \checkmark & \checkmark \\
\cite{Seifert21}      & -- & -- & -- & \checkmark & -- & -- & -- & \checkmark & \checkmark & -- \\
\midrule
\textbf{Ours}         & \checkmark & $\circ$ & \checkmark & \checkmark & \checkmark & $\circ$ & -- & $\circ$ & \checkmark & \checkmark \\
\bottomrule
\end{tabular}

\vspace{2pt}
\parbox{\columnwidth}{\scriptsize MR: multi-radar; MT: multi-person tracking; CH: continuous HAR; ML: ML/DL; AR: activity recognition; FD: fall detection; AD: anomaly detection; LT: long-term monitoring; CS: clinician support; IF: integrated framework. \checkmark: yes; $\circ$: partial; --: no.}
\end{table}

\subsection{Research Gap}

Table~\ref{tab:comparison} compares the reviewed studies across ten
capabilities, from multi-radar sensing and multi-person tracking to
anomaly detection and clinician support, and shows that none of them
covers all of these together. Most studies address one task, such as activity
classification, fall detection, or tracking
\cite{Ullmann23, Pearce23, Hu24}. Systems that come closer to full
integration either rely on a single radar or on non-radar sensors
\cite{Abedi23, Shahid22}, or stop at tracking and fall detection
without activity analysis or a nurse interface \cite{Shen24}. This
paper presents a preliminary study toward closing that gap, guided by
three research questions:
 \begin{itemize}
\item[RQ1:] To what extent can point clouds from multiple
ceiling-mounted radar sensors be fused to provide continuous tracking
of a person across a monitored room?
\item[RQ2:] Can features derived from fused radar point clouds
discriminate between daily activities performed within a continuous,
unsegmented scenario?
\item[RQ3:] How can radar-derived activity information be delivered to
nurses in real time to support assisted-living care?
\end{itemize}

\section{System Architecture}

\begin{figure}[htbp]
    \centering    
    \includegraphics[width=0.80\linewidth]{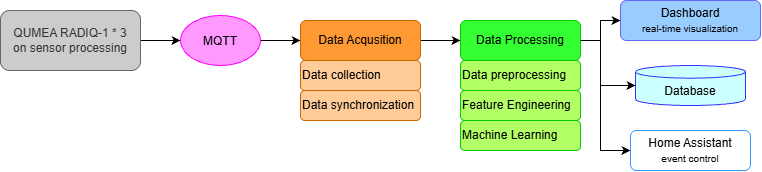}    
    \caption{System architecture of the proposed framework.}    
    \label{fig:system_architecture}
\end{figure}

The system monitors a single room using three ceiling-mounted \href{https://www.qumea.com}{QUMEA} RADIQ-1 radar sensors and has three stages: acquisition, processing and visualization (Fig.~\ref{fig:system_architecture}). Each sensor streams point cloud data over Message Queuing Telemetry Transport (MQTT) to processing pipeline, where the data is fused, tracked, and converted into features for a real-time dashboard. All processing beyond the raw sensor output was developed by the authors for research purposes, therefore the reported results do not reflect the performance of QUMEA products.
\subsection{Data Acquisition}  
\begin{figure}[t]    
    \centering    
    \includegraphics[width=0.75\linewidth]{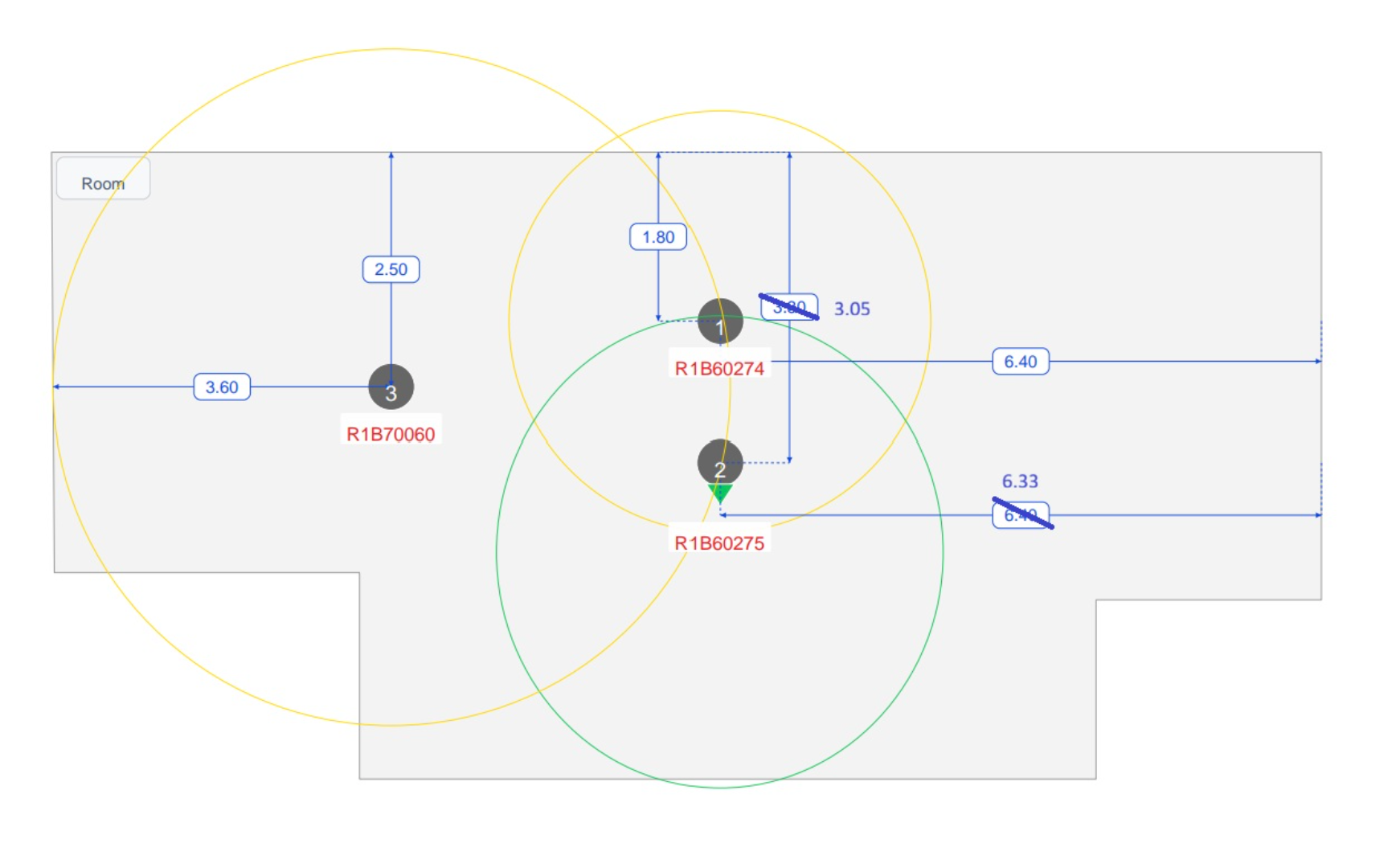}    
    \caption{Layout of the monitored room with the three    ceiling-mounted radar sensors.}    
    \label{fig:room}
\end{figure}

The sensors are mounted in a game room at Satakunta University of Applied Sciences (Fig.~\ref{fig:room}), each covering a radius of about 6.41 m. One sensor is fixed directly to the ceiling at about 5 m, while the other two are suspended below it at about 4 m. One suspended sensor is tilted about 10$^\circ$ because a roof beam prevented a vertical mount, and this tilt also extends coverage to the entrance. Each sensor outputs a three-dimensional point cloud with an SNR value for every point and a zero-Doppler range profile. The sensors publish these directly over authenticated MQTT topics. A Raspberry Pi subscribes to the three topics, groups measurements that arrive within the same
0.5~s window, and logs every processed frame to daily JSONL files. A watchdog thread detects silent MQTT disconnections and reconnects automatically, which keeps long recordings intact.

\subsection{Data Processing}
 \subsubsection{Signal Processing and Sensor Fusion}   

Each point is first transformed from sensor to room coordinates using the sensor's position, pitch, and yaw. Points with an SNR below 5 are
discarded, and the remaining points from all three sensors are merged and clustered with DBSCAN (minimum cluster size of two points). Because one person can appear as separate clusters in different sensors, clusters closer than 0.9~m are merged.
The centroid of each cluster gives a person's position, which is smoothed with an exponential moving average. A person who stands still returns few points, so to avoid flickering tracks, the last position is held for 5~s and a person is marked absent only after ten consecutive empty frames. The
tracker supports up to three people.

\subsubsection{Feature Extraction}

For every frame, the system records presence, position, mean point height, posture, point and cluster counts, per-sensor point counts, a
fall-candidate flag, and the zero-Doppler range profiles. Posture (standing, sitting, or lying) is assigned from thresholds on the mean point height. For activity analysis, frames are grouped into 5~s windows, and eight features are computed per window: the mean, standard deviation, and maximum of the point count; the mean and standard deviation of the mean point height; the fraction of frames with a person present; the mean number of clusters; and the trajectory length. 

\subsubsection{Machine Learning Analysis}
During data collection, an event logger records which activity is
being performed, and each frame and each 5~s window is labeled with
the corresponding activity. A random forest (RF) is trained both on
individual frames, using point count, mean height, cluster count, and
the person's $x$ and $y$ position, and on the eight window features.
Both are evaluated with a random 80/20 split and LOPO
cross-validation. Given the size of the dataset, this stage is treated as a
feasibility study rather than a deployed classifier; results are
reported in Section~\ref{sec:results}.

\subsection{Data Visualization and Integration}

\begin{figure}[t]
    \centering
    \includegraphics[width=0.85\columnwidth]{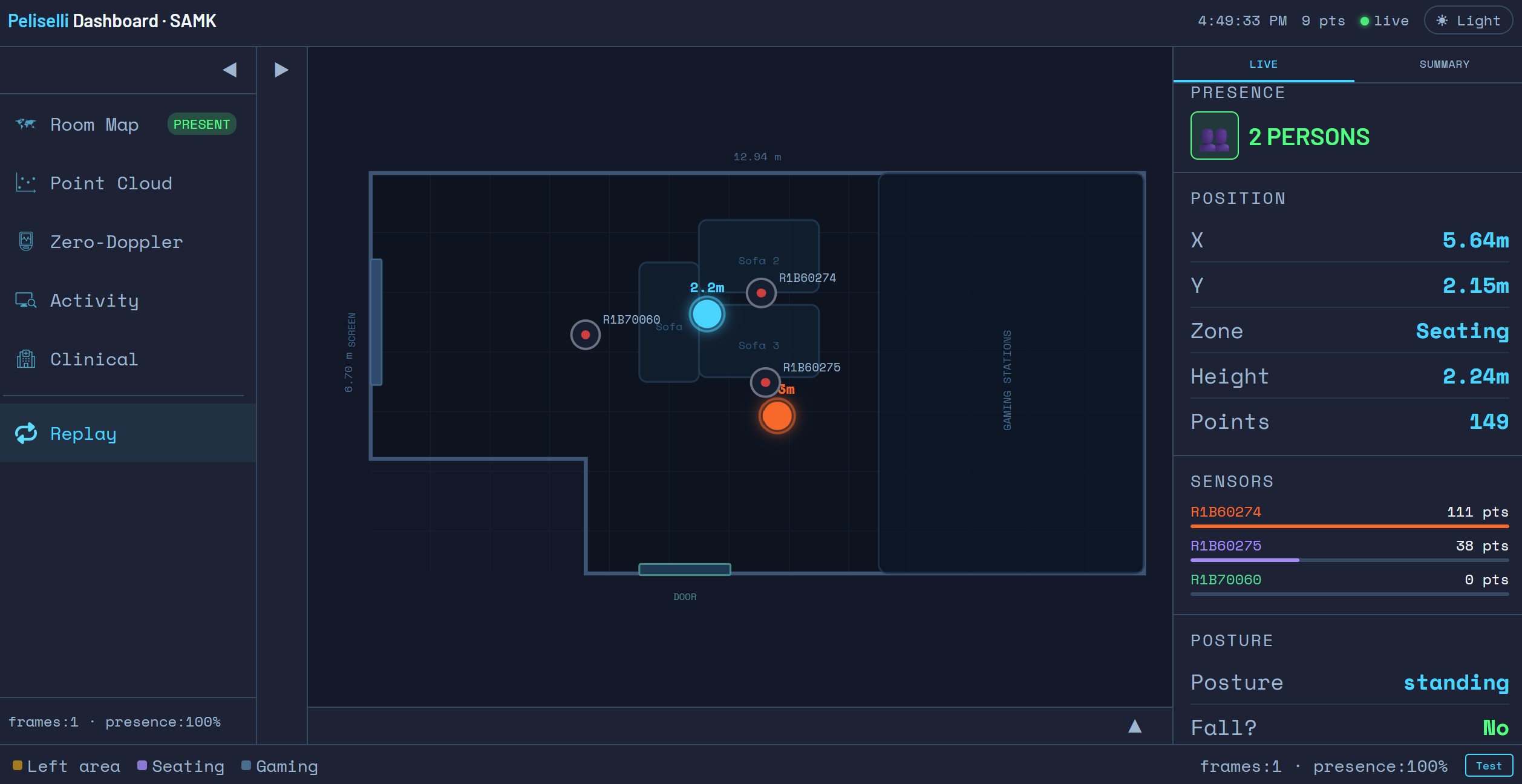}
    \caption{Dashboard showing tracked position of two participants.}
    \label{fig:dashboard}
\end{figure}

\begin{figure}[t]
    \centering
    \includegraphics[width=0.85\columnwidth]{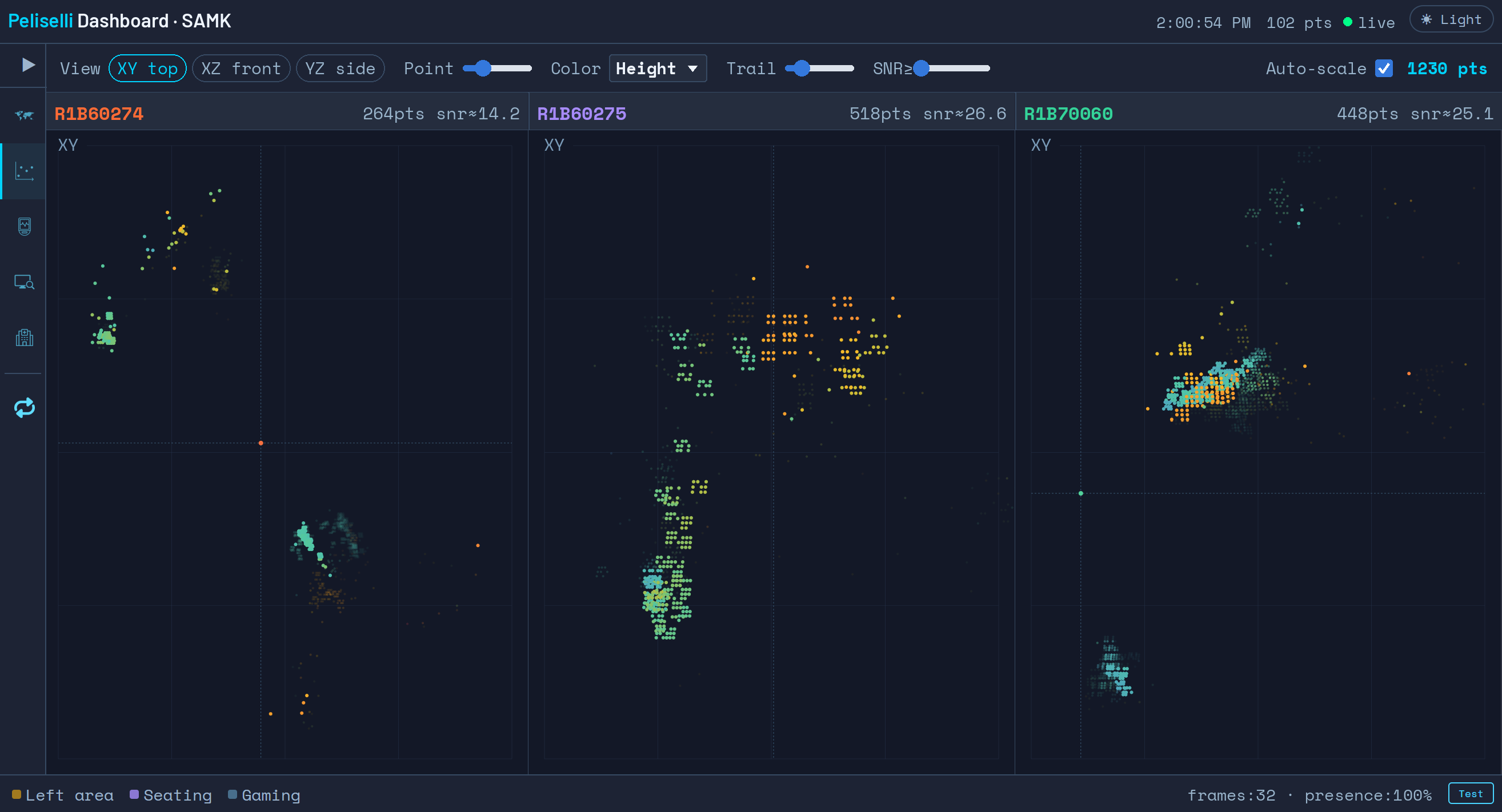}
    \caption{Dashboard showing point cloud of a frame.}
    \label{fig:pointcloud}
\end{figure}

A browser-based dashboard gives nurses a real-time view of the room through five tabs: room map (Fig.~\ref{fig:dashboard}), per-sensor point clouds (Fig.~\ref{fig:pointcloud}), zero-Doppler range profiles, activity timeline, and session replay. In the research setup, events such as a person entering or leaving and posture changes are additionally forwarded via MQTT to Home Assistant for experimental home-automation routines. This configuration is not part of the QUMEA product, has not been validated, and is not intended for care decisions or alarm handling. The same configuration has also been set up in a separate test apartment.
\section{Results}
\label{sec:results}

\subsection{Data Collection}

\begin{figure}[ht]
    \centering
    \includegraphics[width=0.9\columnwidth]{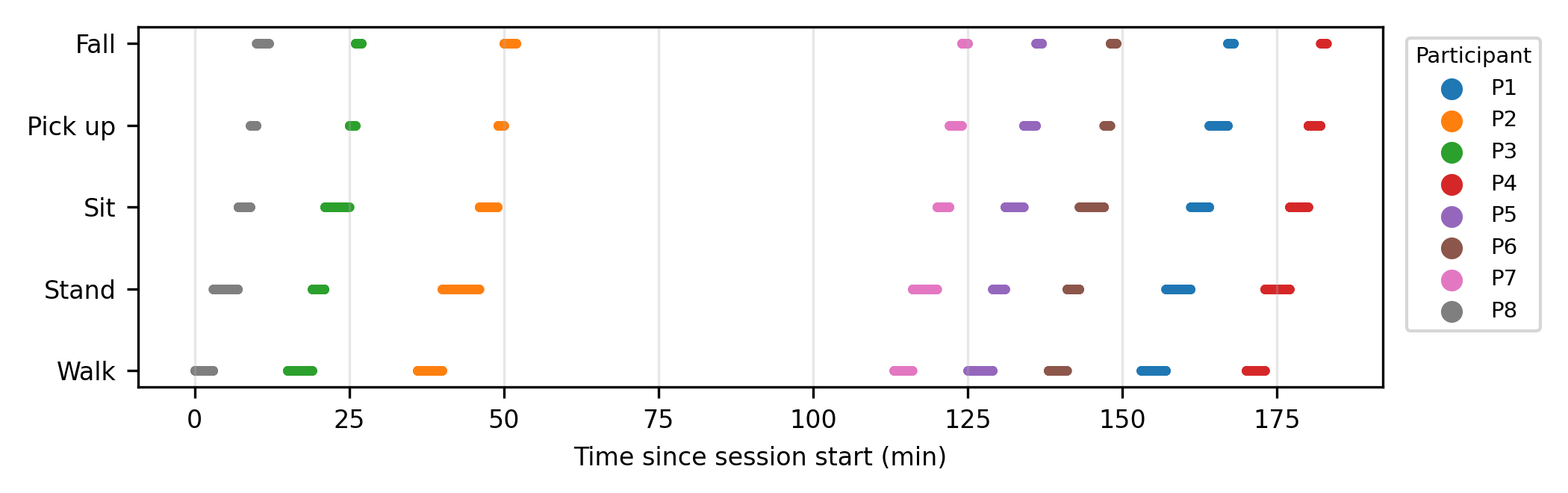}
    \caption{Labeled activities over time for each participant
    (P1--P8), recorded as one continuous scenario.}
    \label{fig:timeline}
\end{figure}

Data were collected in a single session with eight participants. Instead of cueing each activity separately, every participant followed
one continuous scenario containing five target activities: walking, standing, sitting, picking up an object, and falling. Participants
walked between corners of the room while picking up objects, then stood reading a note on the table, and so on, so that activities followed one another with natural transitions, as in daily life
\cite{Ullmann23}. Picking up an object from the ground was kept as a separate class
because, seen from the ceiling, it resembles the start of a fall. For safety, the fall was performed as a controlled movement from kneeling
to lying on the floor. The start and end of each activity were marked with an event logger and aligned with the radar logs by timestamp
(Fig.~\ref{fig:timeline}). In total, the dataset contains about 81~min of labeled recordings, or 4,080 frames and 971 labeled 5~s windows (walking 1,469/325, standing 802/211, sitting 598/174, picking up 641/141, falling 570/120). Throughout the session, the dashboard showed each
participant's position and posture in real time
(Fig.~\ref{fig:dashboard}); since true positions were not recorded, tracking accuracy was not measured.

\subsection{Activity Classification}

The 80/20 split tests subject-dependent performance and LOPO tests performance on unseen participants. Frame-level accuracy was 74\% and 68\%, and window-level accuracy 67\% and 65\%, respectively; windows are fewer than frames, and those at activity transitions contain more than one activity.

\begin{figure}[t]
    \centering
    \includegraphics[width=0.65\columnwidth]{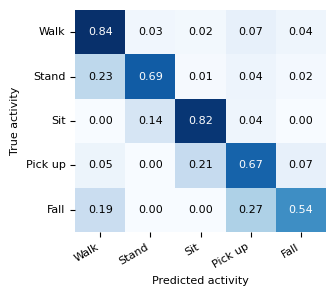}
    \caption{Normalized confusion matrix of the random forest at the
    frame level with a random 80/20 split.}
    \label{fig:cm}
\end{figure}

Fig.~\ref{fig:cm} shows the confusion matrix for the frame-level 80/20
split. Walking was recognized best (recall 0.84), as it is the only
activity with sustained movement across the room. Sitting also reached
a high recall (0.82), while standing (0.69) was most often confused
with walking and sitting, likely because the difference in body height
is small relative to the 5~m sensor height. Participants also fidgeted while standing, and the seated Jenga task involved only small hand movements, producing few radar points. Falling (0.54) and picking up an object (0.67) formed the main confused pair, since both bring
the body close to the floor and the fall was performed slowly. Feature importance was assessed with SHapley Additive exPlanations
(SHAP) \cite{Lundberg17}, using the mean absolute SHAP value over the
test set. Position features dominated, with the person's $y$ and $x$
coordinates together accounting for 68\% of the total, followed by
point count (17\%) and mean height (13\%), while cluster count
contributed little (2\%). This reflects that each activity in the
scenario took place in a typical part of the room, so the classifier
partly learns where an activity happens rather than how the body
moves. Removing the position features reduced frame-level accuracy to 46\% (80/20) and 44\% (LOPO), against 36\% for always predicting the majority class, whereas window-level features, which exclude position, retained most of their accuracy.

\section{Conclusion}

This study demonstrated a multi-radar framework combining tracking, continuous activity analysis, and real-time visualization. Results showed the feasibility of distinguishing five activities with simple point-cloud features, while LOPO results highlighted the challenge of generalizing across participants. Limitations include the small number of participants, the absence of ground-truth positions, and a simulated fall recognized in only about half of the cases, so fall-related performance is not yet established. The 4--5~m mounting height exceeds most patient rooms and may affect results, and accuracy dropped to 46\% without position features, so results may not transfer to other room layouts. Future work will address larger datasets, realistic falls, and typical room configurations.

\section*{Acknowledgment}
The authors thank QUMEA for providing the radar sensors and technical support, and Annariina Lohiranta for her help with data collection. Claude assisted with software development, debugging, and proofreading, and Anara with parts of Section II; the authors reviewed all such content.

\bibliographystyle{IEEEtran}
\bibliography{Bibliography}{}

\end{document}